 \newcommand{\be}[0]{\begin{equation}}
 \newcommand{\ee}[0]{\end{equation}}
 \newcommand{\ba}[0]{\begin{eqnarray}}
 \newcommand{\ea}[0]{\end{eqnarray}}
\documentstyle[12pt, epsfig]{article}
\begin{document}
\Large
\hfill\vbox{\hbox{DTP-95/54}
            \hbox{June 1995}}
\nopagebreak

\vspace{2.75cm}
\begin{center}
\LARGE
{\bf Should the Pomeron and imaginary parts\\ be modelled by
two gluons and real quarks?}
\vspace{0.8cm}
\large

Kirsten B\"uttner

\vspace{0.5cm}
\large
and
\large
\vspace{0.5cm}

M.R. Pennington

\vspace{0.4cm}
\begin{em}
Centre for Particle Theory, University of Durham\\
South Road, Durham, DH1 3LE, U.K.
\end{em}

\vspace{1.7cm}

\end{center}
\normalsize
\vspace{0.45cm}


{\leftskip = 2cm \rightskip = 2cm
We illustrate that solution of the Schwinger-Dyson equation for the
gluon propagator in QCD does not support an infrared softened behaviour, but
only an infrared enhancement. This has consequences for the modelling of the
Pomeron in terms of dressed gluon exchange. It highlights that
an understanding of the Pomeron within QCD must take account of the bound state
nature of hadrons.
\par}
\newpage
\baselineskip=6.7mm
\parskip=2mm

It has long been understood that at high energies total cross-sections for
hadronic processes are controlled by cross-channel Pomeron exchange
\cite{pomeron,Low},
where the Pomeron is believed to be a colour singlet with vacuum quantum
numbers. Low and Nussinov~\cite{Low} proposed a QCD-inspired model
for the Pomeron in terms of two gluon exchange and Landshoff and Nachtmann
\cite{LN}
set up an explicit framework for
phenomenological calculations of the resulting cross-sections. A key
requirement
of their model is that the dressed gluon propagator, $\Delta(k^2)$, should not
have the singularity of the bare massless boson $\sim 1/k^2 \ \mbox{ as }\  k^2
\rightarrow 0$, but should be softened so that the integral
\[ \int^{\infty}_{0} dk^2 \Delta(k^2)^2 \] is finite, where $k$ is a Euclidean
loop momentum.
 Here we discuss whether
such infrared behaviour of the gluon propagator is possible in non-perturbative
QCD.

The infrared behaviour of the gluon propagator is naturally studied in the
continuum using the Schwinger-Dyson equations. It has been known since the work
of Mandelstam~\cite{Mandel} and Bar-Gadda \cite{Bar} that an infrared
enhanced gluon propagator, typically $\Delta(k^2) \sim 1/k^4$, is a possible
solution of the truncated Schwinger-Dyson equations. Baker, Ball and
Zachariasen (BBZ) \cite{BBZ} also deduced such behaviour in axial gauges.
However, their result depends crucially on setting one of the two axial gauge
gluon renormalization functions to zero. West \cite{West} has proved that in
axial gauges, in which only positive norm states occur, a behaviour more
singular than $1/k^2$ is not possible and consequently the neglected axial
gauge
renormalization function must cancel any $1/k^4$ singularity in the infrared.
More recently Cudell and Ross \cite{CR} have shown that an alternative axial
gauge
solution with an infrared softened gluon propagator exists to Schoenmaker's
approximation \cite{Schoen} to the BBZ equation. Unfortunately, this solution
has now been recognised as only having been possible because of an incorrect
sign in Schoenmaker's approximate equation \cite{physrev}.

Because of the difficulty in justifying the neglect of one of the key gluon
renormalization functions in axial gauges, we turn our attention to covariant
gauges and the Landau gauge in particular. In such a gauge $\Delta(k^2) \sim
1/k^4$ has already been shown to be the behaviour of a self-consistent
solution to the gluon Schwinger-Dyson equation~\cite{Mandel,Bar} --- see
\cite{BP} for a full
discussion of the approximations used. Such a $1/k^4$ solution, West
\cite{West2} has argued leads inexorably to a Wilson area law, which many would
regard as a proof of quark confinement. However, such an infrared enhanced
gluon
is at variance with the Landshoff and Nachtmann picture of the Pomeron.
Consequently we should search for an alternative {\it softened} solution to the
Landau gauge gluon equation. We show, here, that no such behaviour is possible.

\baselineskip=7.2mm
To do this we consider the Schwinger-Dyson equation for the gluon propagator in
the Landau gauge. The gluon propagator
is then represented by
\be
\Delta^{\mu\nu}(k)\,=\, \Delta(k^2)\,\left(\delta^{\mu\nu}\,-\,
{k^{\mu}k^{\nu}\over k^2}\right)
\ee
where $ \Delta(k^2)\,=\,G(k^2)/k^2$ with $G(k^2)$ the gluon
renormalization function.
Since confinement must be a result of the non-Abelian nature of QCD, we
consider a
world without quarks. The gluon
Schwinger-Dyson equation may be approximated by treating the ghosts
perturbatively and neglecting 4-gluon interactions, as discussed by Mandelstam
\cite{Mandel}, and replacing
the 3-gluon vertex by its longitudinal component determined by the
Slavnov-Taylor
identity \cite{BP}. The resulting truncated Schwinger-Dyson equation is
displayed
in Fig.~1.

\begin{figure}[h]
\begin{center}
\mbox{\epsfig{file=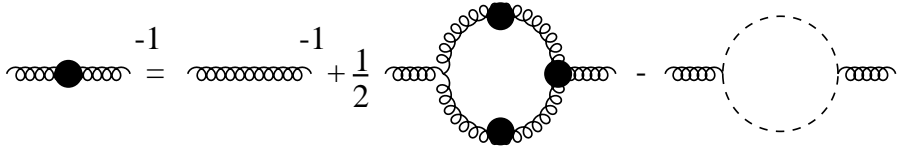,width=13cm}}
\end{center}
\caption[qcd1]{Approximate Schwinger-Dyson equation for the gluon propagator}
\end{figure}

\noindent Then $G(k^2)$ satisfies the following equation:
\ba
\lefteqn{\frac{1}{G(k^2)} =
1 + \frac{g^{2}C_{A}}{96\pi^4}\frac{1}{k^2}\int d^4q \left[ G(q'^2)
A(q^2,k^2) + \frac{G(q^2)G(q'^2)}{G(k^2)}\ B(q^2,k^2)\right.} \nonumber \\
& & \qquad \quad + \left.\frac{G(q^2)-G(k^2)}
{q^2-k^2}\frac{G(q'^2)}{G(k^2)}\ C(q^2,k^2) + \frac{G(q'^2)-G(q^2)}
{q'^2-q^2} \ D(q^2,k^2) \right] ,
\ea
\noindent with $\ q'\ =\ k - q\ $ and where
\ba
  A(q^2,k^2) &=& 48\frac{(q\cdot q')^2}{q^2k^2q'^2} - 64\frac{(q\cdot q')}
{q^2q'^2} + 16\frac{(q\cdot q')^3}{q^2k^2q'^4} - 12\frac{1}{q^2} + 22
\frac{k^2}{q^2q'^2} - 42\frac{(q\cdot q')^2}{q^2q'^4}
- 10\frac{k^4}{q^2q'^4} \nonumber \\
& & + 36\frac{k^2(q\cdot q')}{q^2q'^4}\qquad , \nonumber \\
B(q^2,k^2) &=& - 13\frac{k^2}{q'^4} + 18\frac{k^2(q\cdot q')}{q^2q'^4}
- 2\frac{(q\cdot q')^2}{q^2q'^4} - 4\frac{k^4}{q^2q'^4} + \frac{k^2
(q\cdot q')^2}{q^4q'^4}\; , \nonumber \\
C(q^2,k^2) &=& 4\frac{(q\cdot q')^2}{q^2q'^2} + 6\frac{k^2(q\cdot q')}
{q^2q'^2} + 6\frac{(q\cdot q')}{q'^2} + 8\frac{k^2}{q'^2}
\quad ,  \nonumber \\
D(q^2,k^2) &=& 12\frac{q^2}{q'^2} - 48\frac{(q\cdot q')^2}{k^2q'^2} + 48
\frac{(q\cdot q')^3}{q^2k^2q'^2} +24\frac{(q\cdot q')}{q'^2} -5\frac{k^2}
{q'^2} - 40\frac{(q\cdot q')^2}{q^2q'^2} +9\frac{k^2(q\cdot q')}{q^2q'^2}
\, .\nonumber
\ea
Further approximating $G(q'^2)$ by $G(k^2+q^2)$, which should
 be exact in the infrared limit,
as first proposed by
Schoenmaker \cite{Schoen}, allows
the angular integrals to be performed analytically, giving:
\ba
\lefteqn{\frac{1}{G(k^2)} = 1 + } \nonumber \\
& &\frac{g^{2} C_{A}}{48\pi^2}\frac{1}{k^2}\left\{
\int_{0}^{k^2} dq^2 \left[ G_{1} \left( -1 - 10\frac{q^2}{k^2}
+6\frac{q^4}{k^4} + \frac{q^2}{k^2-q^2} \left( \frac{75}{4} -
\frac{39}{4}\frac{q^2}{k^2} + 4\frac{q^4}{k^4} - 5\frac{k^2}{q^2}\right)
\right) \right.\right. \nonumber \\
& & \left. \qquad \qquad + G_2 \left(-\frac{21}{4}\frac{q^2}{k^2} + 7
\frac{q^4}{k^4} - 3 \frac{q^6}{k^6}\right) +
G_3 \left( \frac{q^2}{k^2-q^2}\left( -\frac{27}{8} - \frac{11}{4}
\frac{q^2}{k^2} - \frac{15}{8}\frac{k^2}{q^2} \right)\right)\right]
\nonumber \\
& & \qquad \qquad +\int_{k^2}^{\infty} dq^2 \left[ G_{1} \left(
\frac{k^2}{q^2} - 6 +
\frac{k^2}{k^2-q^2}\left( \frac{29}{4} + \frac{3}{4}\frac{k^2}{q^2}
\right)\right) \,+ \right.\nonumber \\
& & \qquad \qquad \left.\left. + G_2 \left( - \frac{3}{2} + \frac{1}{4}
\frac{k^2}{q^2} \right) +
G_3 \left( \frac{k^2}{k^2-q^2}\left( \frac{3}{4} -
\frac{67}{8}\frac{k^2}{q^2} - \frac{3}{8}\frac{k^4}{q^4} \right)\right)
\right] \right\}\, ,
\ea
where
\ba
G_{1} &=& G(k^2+q^2) \qquad\qquad ,\nonumber \\
G_{2} &=&  G(k^2+q^2)- G(q^2) \quad ,\nonumber \\
G_{3} &=& \frac {G(q^2)G(k^2+q^2)}{G(k^2)} \; . \nonumber
\ea

In general, this equation has a quadratic ultraviolet divergence, which
would give a mass to the gluon. Such terms have to be subtracted to ensure
the masslessness condition
\be \lim_{k^2 \rightarrow 0} {1\over \Delta(k^2)} = 0 \ \ \mbox{, i.e.}\ \ \
\frac{k^2}{G(k^2)} = 0 \ \ \mbox{for}\ \ k^2 \rightarrow 0 \qquad ,\ee
is satisfied. This property can be derived generally from the Slavnov-Taylor
identity and always has to hold. To determine possible self-consistent
behaviour for the
gluon renormalization function, $G(k^2)$ is expanded in a series in powers of
$k^2/\mu^2$
for $k^2 < \mu^2$ (including possible negative powers).
Here $\mu^2$ is the mass scale above
which we assume perturbation theory applies and  we demand that for  $k^2 >
\mu^2$
the solution of the integral equation matches the perturbative
result, i.e. we have $G(k^2) = 1$ modulo
logarithms.
To check whether Eq.~(2) allows an infrared softened gluon propator, i.e. the
gluon renormalization function to vanish in the infrared, we take (cf.
\cite{CR})
\be
G_{in}(k^2) = \left\{ \begin{array}{ll}
\left({k^2}/{\mu^2}\right)^{1-c} & \mbox{if $k^2< \mu^2$} \\ \\
1 & \mbox{if $k^2> \mu^2$}
\end{array}
\right. \ee
\noindent as a trial input function and substitute it into the right hand side
of the
integral equation, Eq.~(2). $c$ in Eq.~(5) has to be positive to satisfy
Eq.~(4).
Performing the $q^2$-integration, we obtain, after mass renormalization:
\ba
\frac{1}{G_{out}(k^2)} &=& 1 + \frac{g^2C_{A}}{48\pi^2} \left[
\ D_{1} + \ D_{2} \left( \frac{\mu^2}{k^2}\right)^{1-c}
+ \ D_{3} \left(\frac{k^2}{\mu^2} \right)^{1-c} +\  D_{4}\left(
\frac{k^2}{\mu^2}\right)^{c} + ...\right]  \ ,
\ea
\noindent where $G_{1}, G_{2}$ and $G_{3}$ have been expanded
for small $k^2$ and only the first few terms have been collected in this
equation so that
\ba
D_{1} &=& -\left( \frac{3}{2} + \frac{5+6c}{1-c} +\frac{25}{4}\ln\left(
\frac{\Lambda^2}{\mu^2} \right) \right)\nonumber \qquad , \\
D_{2} &=& -\left( \frac{3}{4(2-2c)} +\frac{3}{4}\ln\left(
\frac{\Lambda^2}{\mu^2} \right) \right)\nonumber \qquad , \\
D_{3} &=& -\frac{1971}{60} +\frac{29c}{2} +\frac{37}{20c} +
\frac{6-13c}{2(1-c)} +\frac{59-32c}{4(2-c)}
+\frac{155-64c}{8(3-c)} \nonumber \\
& &+\frac{127-49c}{8(4-c)} +\frac{23-11c}{4(5-c)} -\frac{125+61c}{8(1-2c)}
-\frac{55+6c}{8(2-2c)} +\frac{3}{4(3-2c)} \nonumber \\
& &-8(2-c)\Psi(-2c) -8(2-c)\Psi(1) \phantom{\frac{2}{2}}\nonumber
\qquad , \\
D_{4} &=& \frac{61+6c}{8(1-2c)}\nonumber \qquad ,
\ea
\noindent where $\Psi$ is the logarithmic derivative of the Gamma
function.
Thus the dominant infrared behaviour is:
\be
\frac{1}{G_{out}(k^2)} \rightarrow - \left(\frac{\mu^2}{k^2}
\right)^{1-c}
\ee
and self-consistency is spoiled by a negative sign, just as in axial gauges
\cite{physrev}. Note that higher order terms in $k^2$ in the input form of
Eq.~(5) have no qualitative effect.
We thus see that an infrared softened gluon is not possible. Even softer gluons
resulting from the dynamical generation of a gluon mass, though often claimed,
only arise if multi-gluon vertices have massless particle singularities that
stop the zero momentum limit of the Slavnov-Taylor identity being smooth. Such
singularities, though they occur in the vertices of Stingl et al.
\cite{Stingl},
should not be present in QCD. In contrast, similar arguments to the above show
 that an infrared enhanced behaviour
$G(k^2) \sim 1/k^2$ for $k^2 \to 0$ is a consistent solution~\cite{physrev}.

This gives the  confining gluon behaviour of $\Delta(k^2) \sim 1/k^4$.
Such a gluon has no Lehman spectral representation \cite{SDE} and so is not
a physical state, but is confined. How does this infrared behaviour of the
gluon affect the Pomeron of Landshoff and Nachtmann? Their belief in an
infrared
softened, rather than enhanced, gluon rests on their model requirement that the
integral \[ \int^{\infty}_{0} dk^2 \Delta(k^2)^2 \] should be finite. However,
as we now explain we do not believe the issue of whether this integral is
finite
or not is relevant to the finiteness of total cross-sections. The
Landshoff-Nachtmann picture is to imagine that the two dressed gluons that
model
their Pomeron couple to single quarks with other quarks in each initial state
hadron being spectators (Fig.~2a).

\begin{figure}[th]
\begin{center}
\mbox{\epsfig{file=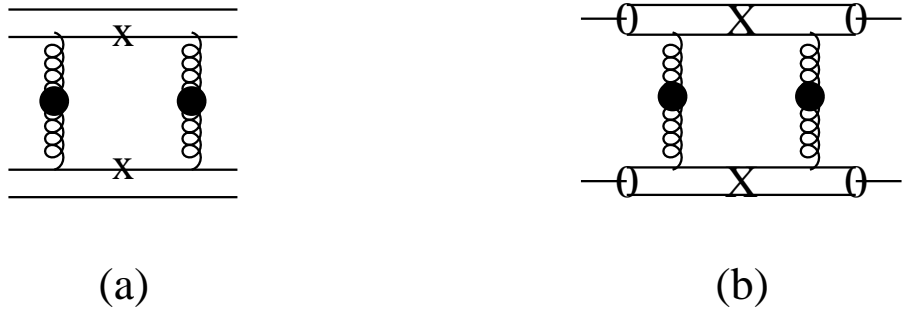,width=13cm}}
\end{center}
\caption[pomeron]{Diagrammatic representation of the Pomeron in meson-meson
scattering:\\
(The lines marked with an $X$ are on-shell in the determination of the total
cross-section.)\\
 (a) Exchange of a gluon pair between two quarks (Landshoff-Nachtmann model),\\
 (b) Exchange of a gluon pair between two hadrons.}
\end{figure}

In this way the forward hadronic scattering amplitude is viewed as essentially
quark-quark scattering (Fig.~2a). The total cross-section is then just the
imaginary part of this forward elastic quark scattering amplitude, by the
optical theorem. However, an imaginary part is only generated if the quarks can
be on mass-shell and have poles in their propagators, as an electron or pion
does. This assumption is the key to the Landshoff-Nachtmann picture (Fig.~2a)
and the subsequent phenomenology. However, quarks are confined particles; their
propagators are likely entire functions and the elastic quark amplitude has no
imaginary part. Only an infrared enhanced gluon propagator has been shown to
produce a confined light quark propagator \cite{SDE}. It is then the bound
state
properties of hadrons that are the essential ingredients of total
cross-sections. It is the intermediate hadrons that have to be on-shell
(Fig.~2b)
and not the confined quarks. Confinement requires that hadronic amplitudes are
not merely the result of free quark interactions. Only for hard short distance
processes is such a perturbative treatment valid. In soft physics, the bound
state nature of light hadrons has to be solved to compute observables.
A programme of research to solve the appropriate Bethe-Salpeter equations is
under way~\cite{ANU}. Indeed, no processes
are softer than those that produce hadronic total cross-sections. Consequently,
an infrared enhanced gluon propagator is not at variance with the Pomeron, but
is in fact in accord with quark confinement and with
low energy properties of hadrons like dynamical chiral symmetry breaking

\vspace{0.6cm}
{\bf {\large Acknowledgements}}
\vspace{0.4cm}

KB thanks the University of Durham for the award of a research studentship.
We are grateful to the National Centre for Theoretical Physics at the
Australian National
University for their hospitality, where this work was carried out.


\begin{thebibliography}{99}
\baselineskip=6mm
\bibitem{pomeron} P.D.B. Collins, \lq\lq  Regge theory and high energy
physics",
Cambridge:\\ Cambridge University Press 1977.
\bibitem{Low} F.E. Low, Phys. Rev. D12 (1975) 163;\\
S. Nussinov, Phys. Rev. Lett. 34 (1975) 1286.
\bibitem{LN} P.V. Landshoff and O. Nachtmann, Z. Phys. C39 (1989) 405.
\bibitem{Mandel} S. Mandelstam, Phys. Rev. D20 (1979) 3223.
\bibitem{Bar} U. Bar-Gadda, Nucl. Phys. B163 (1980) 812.
\bibitem{BBZ} M. Baker, J.S. Ball and F. Zachariasen, Nucl. Phys. B186 (1981)
531.
\bibitem{West} G.B. West, Phys. Lett. B115 (1982) 468.
\bibitem{CR} J.R. Cudell and D.A. Ross, Nucl. Phys. B359 (1991) 247.
\bibitem{Schoen} W.J. Schoenmaker, Nucl. Phys. B194 (1982) 535.
\bibitem{physrev} K. B\"uttner and M.R. Pennington,
Univ. of Durham preprint DTP-95/32 (June 1995).
\bibitem{BP} N. Brown and M.R. Pennington, Phys. Rev. D39 (1989) 2723.
\bibitem{West2} G.B. West, Phys. Rev. D27 (1983) 1878.
\bibitem{Stingl} U.H\"abel, R. K\"onning, H.G. Reusch, M. Stingl and
S. Wigard,\\
Z.Phys. A336 (1990) 435.
\bibitem{SDE} C.D. Roberts and A.G. Williams,\\
\lq\lq Dyson-Schwinger Equations and their Application to Hadronic Physics", \\
 Progress in Particle and Nuclear Physics 33 (1994) 477.
\bibitem{ANU} A. Bender, C.J. Burden, R.T. Cahill, C.D. Roberts,
 P.C. Tandy and M.J. Thomson,\\
contributions to the Workshop on {\it Non-Perturbative Methods in Field
Theory},\\ Canberra (May, 1995).
\end{thebibliography}
\end{document}